\documentclass[12pt]{elsart}

\usepackage{amsmath}
\usepackage{amsfonts}
\usepackage{graphicx}

\begin{document}

\begin{frontmatter}

\title{Statistical complexity, Fisher-Shannon information, and Bohr orbits in the $H$-atom}

\author[jsr]{Jaime Sa\~{n}udo} and
\ead{jsr@unex.es}
\author[rlr]{Ricardo L\'{o}pez-Ruiz}
\ead{rilopez@unizar.es}

\address[jsr]{
Departamento de F\'isica, Facultad de Ciencias, \\
Universidad de Extremadura, E-06071 Badajoz, Spain}

\address[rlr]{
DIIS and BIFI, Facultad de Ciencias, \\
Universidad de Zaragoza, E-50009 Zaragoza, Spain}


\begin{abstract}
The Fisher-Shannon information and a statistical measure of complexity 
are calculated in the position and momentum spaces for the wave functions of the $H$-atom. 
For each level of energy, it is found that these two indicators take their 
minimum values on the orbitals that correspond to the classical (circular) orbits in the Bohr 
atomic model, just those with the highest orbital angular momentum. 
\end{abstract}


\end{frontmatter}

\maketitle

The atom can be considered a complex system. Its structure can be determined 
through the well established equations of Quantum Mechanics \cite{landau1981,galindo1991}. 
Depending on the set of quantum numbers defining the state of the atom, different 
conformations are avalaible to it. As a consequence, if the wave function of the atomic state 
is known, the probability densities in the position and the momentum spaces can be obtained.
In this case, the tools that have been developed
in information and complexity theories in the last years can be applied \cite{gadre1987,panos2005}.  
Among them, several statistical magnitudes such as Shannon and Fisher informations,
different indicators of complexity, etc., can be 
calculated with a low computational cost 
\cite{gadre1985,dehesa1994,dehesa2004,dehesa2005,panos2007,sen2007,sen2008}.

These quantities enlighten new details of the hierarchical organization of the atomic states.
In fact, states with the same energy can display, for instance, different values of complexity.
Then, it could be interesting to study if the hierarchy introduced by these statistical magnitudes
could have some physical interpretation.  It is our aim to follow this insight.
Thus, in this letter, we test this possibility by performing the calculations 
of some of these magnitudes on the simplest atomic system,
that is, the hydrogen atom ($H$-atom).

Let us start by recalling the non-relativistic wave functions of the $H$-atom.
These functions in position space ($\vec{r}=(r,\Omega)$, with $r$ the radial distance and
$\Omega$ the solid angle) are:
\begin{equation}
\Psi_{n,l,m}(\vec{r})= R_{n,l}(r)\; Y_{l,m}(\Omega)\;,
\end{equation}
where $R_{n,l}(r)$ is the radial part and $Y_{l,m}(\Omega)$ is the spherical harmonic 
of the atomic state determined by the quantum numbers $(n,l,m)$. The radial part is expressed 
as \cite{galindo1991}
\begin{equation}
R_{n,l}(r)= {2\over n^2} \left[{(n-l-1)!\over (n+l)!}\right]^{1/2}\;
\left({2r\over n}\right)^l\;e^{-{r\over n}}\; L_{n-l-1}^{2l+1}\left({2r\over n}\right)\;,
\end{equation}
being $L_{\alpha}^{\beta}(t)$ the associated Laguerre polynomials.
Atomic units are used through the text.

The same functions in momentum space ($\vec{p}=(p,\hat{\Omega})$, 
with $p$ the momentum modulus and $\hat{\Omega}$ the solid angle) are:
\begin{equation}
\hat{\Psi}_{n,l,m}(\vec{p})= \hat{R}_{n,l}(p)\; Y_{l,m}(\hat{\Omega})\;,
\end{equation}
where the radial part $\hat{R}_{n,l}(p)$ is now given by the expression \cite{bethe1977}
\begin{equation}
\hat{R}_{n,l}(p)= \left[{2\over\pi}{(n-l-1)!\over (n+l)!}\right]^{1/2}\;
n^2\;2^{2l+2}\;l!\;{n^lp^l\over (n^2p^2+1)^{l+2}}\; 
C_{n-l-1}^{l+1}\left({n^2p^2-1\over n^2p^2+1}\right)\;,
\end{equation}
with $C_{\alpha}^{\beta}(t)$ the Gegenbauer polynomials.

Taking the former expressions, the probability density
in position and momentum spaces,
\begin{equation}
\rho(\vec{r})\;=\;\mid\Psi_{n,l,m}(\vec{r})\mid^2\;, \hspace{1cm}
\gamma(\vec{p})\;=\;\mid\hat{\Psi}_{n,l,m}(\vec{p})\mid^2\;,
\end{equation}
can be explicitly calculated. From these densities, we proceed now to compute
the statistical complexity and the Fisher-Shannon information.

First, the measure of complexity $C$ recently introduced by Lopez-Ruiz, Mancini and Calbet
\cite{lopez1995,lopez2001,lopez2002,lopez2005},
the so-called $LMC$ complexity, is defined as
\begin{equation}
C = H\cdot D\;,
\end{equation}
where $H$ represents the information content of the system and $D$ is the
distance from the actual state of the system to some prestablished reference state. 

For our purpose, we take a version used in Ref. \cite{lopez2002}
as quantifier of $H$. This is the simple exponential Shannon entropy,
that in the position and momentum spaces takes the form, respectively,
\begin{equation}
H_r = e^{S_r}\;, \hspace{1cm}
H_p = e^{S_p}\;.
\end{equation}
$S_r$ and $S_p$ are the Shannon information entropies \cite{shannon1948},
\begin{equation}
S_r = -\int\rho(\vec{r})\;\log \rho(\vec{r})\; d\vec{r}\;, \hspace{1cm}
S_p = -\int\gamma(\vec{p})\;\log \gamma(\vec{p})\; d\vec{p}\;.
\end{equation}
We keep for the disequilibrium the form originally introduced in 
Refs. \cite{lopez1995,lopez2002}, that is,
\begin{equation}
D_r = \int\rho^2(\vec{r})\; d\vec{r}\;, \hspace{1cm}
D_p = \int\gamma^2(\vec{p})\; d\vec{p}\;,
\end{equation}
In this manner, the final expressions for $C$ in position and 
momentum spaces are:  
\begin{equation}
C_r = H_r\cdot D_r\;, \hspace{1cm}
C_p = H_p\cdot D_p\;.
\end{equation}
These quantities are plotted in Fig. 1 as function of the modulus of the
third component $m$ of the orbital angular momentum $l$ for different pairs 
of $(n,l)$ values. Let us recall at this point the range of the quantum numbers: 
$n\geq 1$, $0\leq l \leq n-1$, and $-l\leq m \leq l$.
Fig. 1(a) shows $C_r$ for $n=15$ and Fig. 1(b) shows $C_ r$ for $n=30$.
In both figures, it can be observed that $C_r$ splits in different sets of discrete points.
Each one of these sets is associated to a different $l$ value. 
It is worth to note that the set with the minimum values of $C_r$ corresponds just
to the highest $l$, that is, $l=n-1$. 
The same behavior can be observed in Figs. 1(c) and 1(d) for $C_p$.

Second, other types of statistical measures that maintain the product form of $C$ can be defined.
Let us take, for instance, the Fisher-Shannon $P$ information that has been also applied
in Refs. \cite{dehesa2004,sen2008} in atomic systems. This quantity, in the position 
and momentum spaces, is given respectively by   
\begin{equation}
P_r = J_r\cdot I_r\;, \hspace{1cm}
P_p = J_p\cdot I_p\;,
\end{equation}
where the first factor
\begin{equation}
J_r = {1\over 2\pi e}\;e^{2S_r/3}\;, \hspace{1cm}
J_p = {1\over 2\pi e}\;e^{2S_p/3}\;,
\end{equation}
is a version of the exponential Shannon entropy \cite{dembo1991}, 
and the second factor
\begin{equation}
I_r = \int{[\vec{\nabla}\rho(\vec{r})]^2\over \rho(\vec{r})}\; d\vec{r}\;, \hspace{1cm}
I_p = \int{[\vec{\nabla}\gamma(\vec{p})]^2\over \gamma(\vec{p})}\; d\vec{p}\;,
\end{equation}
is the so-called Fisher information measure \cite{fisher1925}, that quantifies the narrowness 
of the probability density. This last indicator
can be analytically obtained in both spaces (position and momentum). The results are \cite{dehesa2005}:
\begin{equation}
I_r={4\over n^2}\;\left(1-{|m|\over n}\right),
\end{equation}
\begin{equation}
I_p={2n^2}\;\left\{5n^2+1-3l(l+1)-(8n-3(2l+1))\;|m|\right\}\;.
\end{equation}

Fig. 2 shows the calculation of $P$ as function of the modulus of the
third component $m$ for different pairs of $(n,l)$ values. 
In Fig. 2(a), $P_r$ is plotted for $n=15$, and $P_ r$ is plotted for $n=30$ in Fig. 2(b).
Here $P_r$ also splits in different sets 
of discrete points, showing a behavior parallel to the above signaled for $C$ (Fig. 1). 
Each one of these sets is also related with a different $l$ value. 
We remark again that the set with the minimum values of $P_r$ corresponds just
to the highest $l$. In Figs. 2(c) and 2(d), the same behavior can be observed for $P_p$.

Let us finish this short note with the conclusions.
The statistical complexity and the Fisher-Shannon information have been computed on the $H$-atom.
In this case, we have taken advantage of the exact knowledge of the wave functions.
Concretely, we put in evidence that, for a fixed level of energy $n$, these statistical magnitudes
take their minimum values for the highest allowed orbital angular momentum, $l=n-1$. 
It is worth to remember at this point that the radial part of this particular wave function,
that describes the electron in the $(n,l=n-1)$ orbital, has no nodes. This means
that the spatial configuration of this atomic state is, in some way, a spherical-like shell. 
In fact, the mean radius of this shell, $<r>_{n,l,m}$, which is given by \cite{eisberg1961}  
\begin{equation}
<r>_{n,l=n-1,m}\equiv <r>_{n,l=n-1} = n^2\left(1+{1\over 2n}\right),
\end{equation}
tends, when $n\gg 1$, to the radius of the $nth$ energy level, 
$r_{Bohr}=n^2$, of the Bohr atom. 

It is remarkable that the minimum values of these statistical measures calculated
from the quantum wave functions of the $H$-atom enhance our intuition by selecting 
just those orbitals that in the pre-quantum image are the Bohr orbits. Therefore,
we conclude that the results here reported show that new insights could 
be inferred from these magnitudes at the quantum level.


\newpage

\begin{figure}[h]
\centerline{\includegraphics[width=7cm]{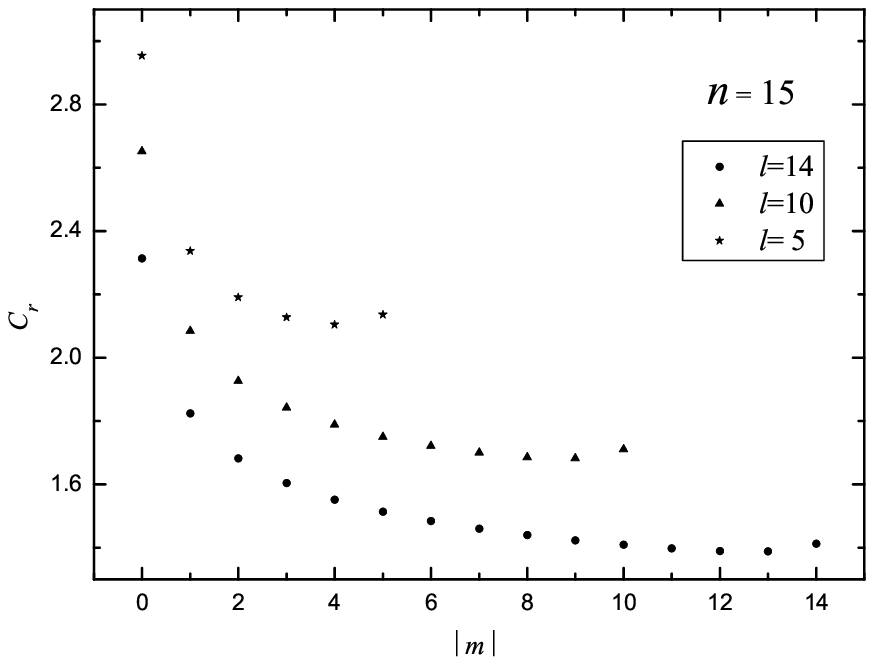}\hskip 5mm\includegraphics[width=7cm]{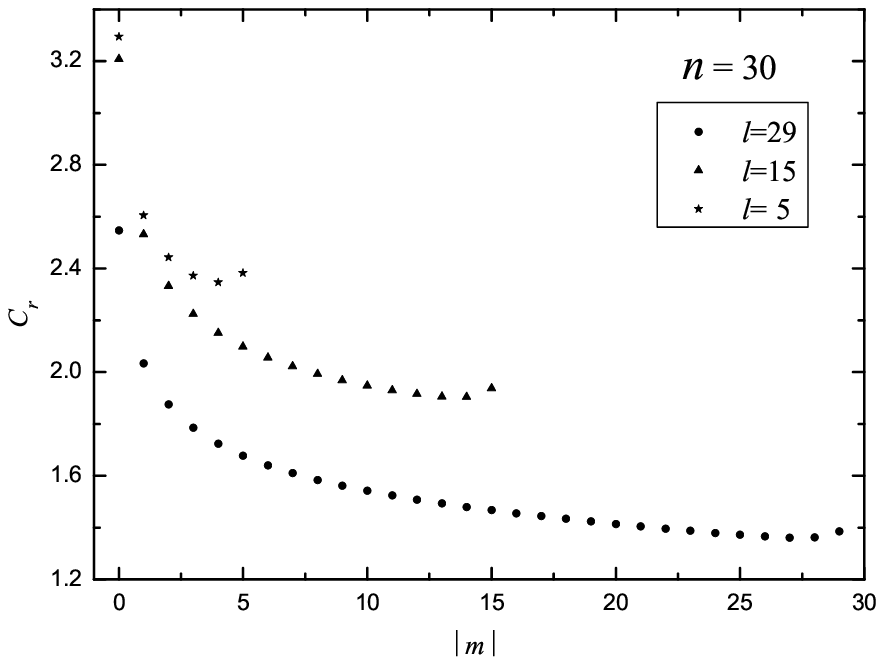}}
\centerline{(a)\hskip 7cm (b)} 
\centerline{\includegraphics[width=7cm]{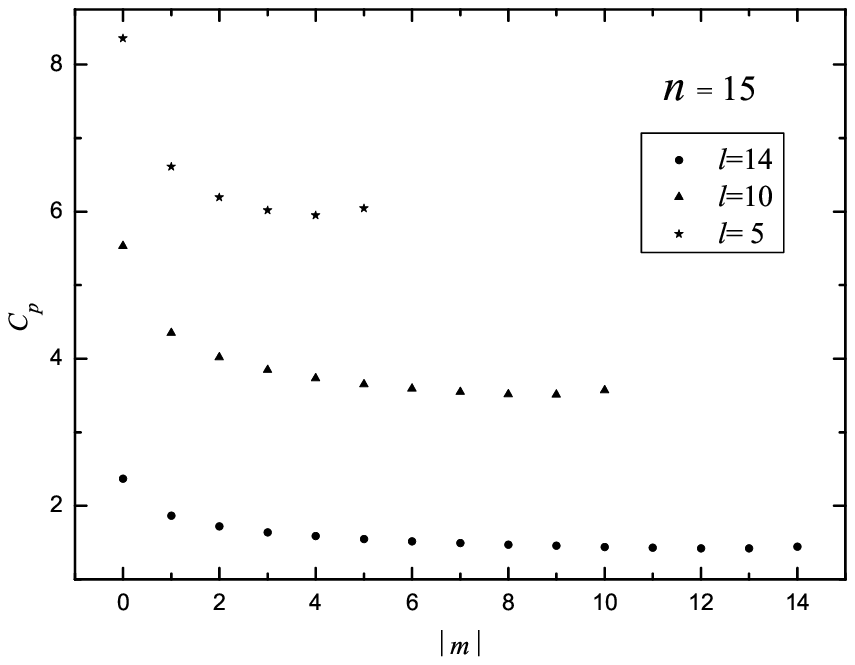}\hskip 5mm\includegraphics[width=7cm]{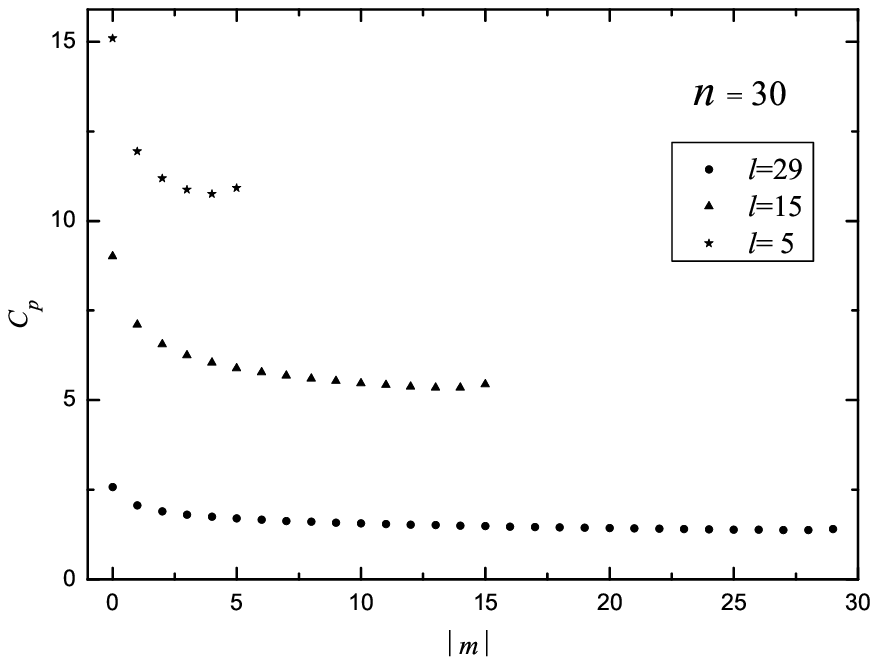}} 
\centerline{(c)\hskip 7cm (d)} 
\caption{Statistical complexity in position space, $C_r$, and momentum space, $C_p$, vs. 
$|m|$ for different $(n,l)$ values in the hydrogen atom. 
$C_r$ for (a) $n=15$ and (b) $n=30$. $C_p$ for (c) $n=15$ and (d) $n=30$. 
All values are in atomic units.}
\label{fig1}
\end{figure}

\newpage
\begin{figure}[h]
\centerline{\includegraphics[width=7cm]{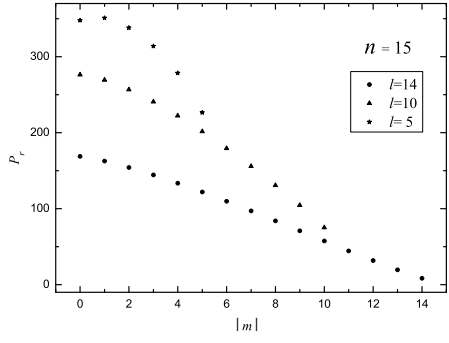}\hskip 5mm\includegraphics[width=7cm]{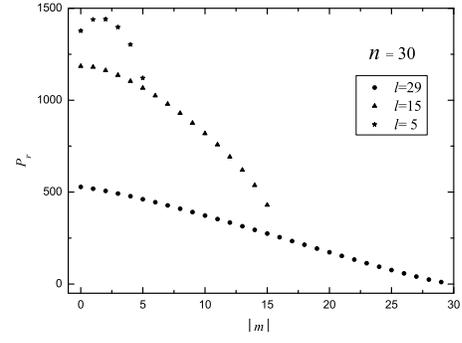}}
\centerline{(a)\hskip 7cm (b)} 
\centerline{\includegraphics[width=7cm]{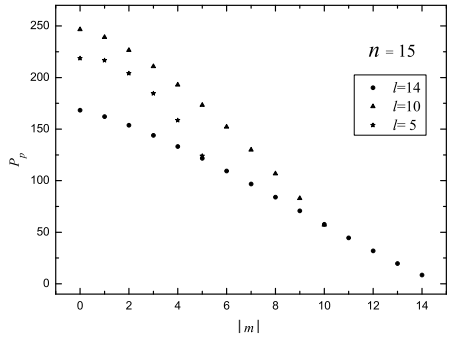}\hskip 5mm\includegraphics[width=7cm]{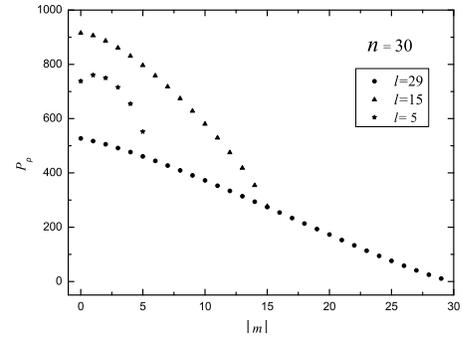}} 
\centerline{(c)\hskip 7cm (d)} 
\caption{Fisher-Shannon information in position space, $P_r$, and momentum space, $P_p$, vs. 
$|m|$ for different $(n,l)$ values in the hydrogen atom. 
$P_r$ for (a) $n=15$ and (b) $n=30$. $P_p$ for (c) $n=15$ and (d) $n=30$. 
All values are in atomic units.} 
\label{fig2}
\end{figure}

\end{document}